\documentclass[12pt,a4]{article}\input{epsf.tex}
\begin{document}
\begin{center}
{\bf LEVEL DENSITY AND RADIATIVE STRENGTH FUNCTIONS IN LIGHT NUCLEI: $^{60}$Co
AS AN EXAMPLE OF THE METHOD FOR DETERMINATION AND THEIR RELIABILITY VERIFICATION\\}
\end{center}
\begin{center} 
{\bf A.M. Sukhovoj,
 V.A. Khitrov}\\
{\it 
  FLNP, Joint Institute for Nuclear
Research, Dubna, Russia}\\
{\bf Pham Dinh Khang}\\
{ \it  National University of Hanoi}\\
{\bf  Vuong Huu Tan, Nguyen Xuan Hai}\\
{\it Vietnam Atomic Energy Commission}\\
\end{center}

\section*{Introduction}
 \hspace*{14pt}

The question about the reliability of observables is very important for
 every experimental work.

Level density and radiative strength functions of cascade gamma-transitions
 below the neutron binding energy $B_n$ are typical example  of two problems 
 which should be solved in order to provide high reliability of the observables.
It is necessary: 

1. to develop appropriate method for extraction of the desired parameters from
 experimental spectra when they cannot be observed directly;

2. to get maximum possible statistics of the experimental data which would
provide uniqueness of the determined parameters and possibility for their
additional independent verification.

Up to now level density $\rho$ was obtained from the experimental data on:

(a) neutron evaporation spectra (see, for example, [1]);

(b) intensities of the two-step gamma-cascades [2]
 \begin{equation}
 I_{\gamma\gamma}=\sum_{\lambda ,f}\sum_{l}\frac{\Gamma_{\lambda l}}
 {\Gamma_{\lambda}}\frac{\Gamma_{lf}}{\Gamma_l}=\sum_{\lambda ,f} 
\frac{\Gamma_{\lambda l}}{<\Gamma_{\lambda l}>
 m_{\lambda l}} n_{\lambda l}\frac{\Gamma_{lf}}{<\Gamma_{lf}> m_{lf}},
\end{equation}
connecting compound state (neutron resonance) with the group of low-lying levels
of the studied nucleus and determined according to [3] for all possible energy
intervals of their primary transitions;

(c) gamma-ray spectra following de-population of levels with the excitation
energy $E_{ex}$ [4] in nuclear reactions like (d,p) and $(^3$He,$\alpha$) [5].

In the methods [2] and [5], $\rho$ is estimated simultaneously with the
radiative strength functions of cascade gamma-transitions

\begin{equation}
k=f \times A^{2/3}=\Gamma_{\lambda l}/(E_{\gamma}^3\times A^{2/3}\times
 D_{\lambda}).
\end{equation}
Here $A$ is the mass of a nucleus, $E_{\gamma}$ is the energy of
gamma-transition, $D_{\lambda}$ is the mean spacing between decaying levels
$\lambda$,  $\Gamma_{\lambda l}$ is the mean partial width of transitions
between the compound state $\lambda$ and some set of levels $l$.

It is obvious that the criterion of reliability of determination of $\rho$ and
$k$ is the agreement between the results of different experiments or, at least,
possibility to explain origins of existing discrepancy.

\section{Main sources of systematical uncertainty in determination of $\rho$
within different methods}
The presence of serious enough discrepancy between level densities determined
in the frameworks of different methods [1,5] and [2] requires one to estimate
reliability of these results.

In methods [1,5], the main problem for extraction of, for example, $\rho$ is
the proportionality of intensity of the registered spectra to  product of
number of levels $m=\rho \times \Delta E$   in a given excitation energy
interval $\Delta E$ and emission probability  $T$ of the reaction products.
In addition to infinite number of functional dependences $\rho=f(E_{ex})$ and
$T=\phi(E_{ex})$ reproducing corresponding spectra, the values of level
density and emission probability of reaction product $T$ can vary in interval
from $-\infty$ to  $+\infty$. I.e., the kind of spectra measured in methods
[1,5] leads to low confidence level of the obtained results.

Besides, the value of $\rho$ is determined in [1] from experimental data with
the use of the model calculated penetrability $T$ of nuclear surface for
evaporated nucleon (light nucleus). It is to be calculated to a precision of
some tens of percent with accounting for possible high-frequent sign-variable
variation of $T$ with respect to its average value. This effect is directly
observed at extraction of $\rho$ and $k$ from intensities of the two-step
cascades.

As a result, the confidence of the results like that given in [1] cannot be
estimated impartially. For example, there are no reasons to exclude the
possibility of compensation  of decrease (increase) in level density in some
excitation energy interval by increase (decrease) in emission probability of
neutron in evaporation spectra or gamma-quantum in the primary transition
spectra depopulating levels in the vicinity of the excitation energy $E_{ex}$. 
Such situation is directly observed [6] for $\rho$ and $k$ derived from the 
intensities $I_{\gamma\gamma}$. Optical potentials used in methods like [1], 
most probably, cannot provide calculation of $T$ in required details and 
guaranteed error in determination of  $\rho$ at the level achieved in [6].

All results in [5] were obtained in the methodical variant containing several
sources of systematical errors that lead to unknown resulting error for both
$\rho$ and $k$. Main part of them, however, can be [7] removed or noticeably 
reduced by suitable modification of the methods of obtaining spectra of 
primary transitions [8] and determination [9] from them of reliable values of 
the parameters $\rho$ and $k$. The difference of energy dependence of the 
radiative strength functions of gamma-transitions with equal energy but 
de-populating levels with different energy $E_{ex}$ must be taken into account
in [5], as well. According to [10], strong difference between strength
functions of primary and secondary transitions manifests itself in different
nuclei from the mass region $40 \le A \le 200$. 

Noticeably more favorable situation is with the achieved confidence of $\rho$
and $k$ values obtained with the use of [2] and possibility to improve it.
It is obvious that level densities and strength functions derived from the 
experimental cascade intensities $I_{\gamma\gamma}$ contain both ordinary 
statistic and specific systematic errors.

Probable value of ordinary errors can be easily obtained from estimation of 
the top uncertainty of the observed cascade intensity distributions by means 
of the standard formula of error transfer. For example, comparison of 
intensities of the cascade primary transitions from [11] and [12] used for 
normalization of $I_{\gamma\gamma}=F(E_1)$ and application of the results of 
extrapolation [13] of distributions of random cascade intensities with the 
energies of their intermediate levels $E_{ex} <0.5B_n$ to the zero detection 
threshold provides realistic estimation of systematic errors of both amplitude
and form of function 
$F(E_1)$.

Variation of its values as input data [2] permits one easily enough and 
reliably to estimate ordinary systematic errors of $\rho$ и $k$ at the 
presents of non-linear relation between $\delta F(E_1)$, $\delta \rho$ and 
$\delta k$.

As it was obtained in [14], $\delta \rho$ and  $\delta k$ estimated in this 
way cannot explain step-like structures in energy dependence of $\rho$ 
determined within  method [2] and discrepancy of these results with 
conventional ideas of ``smooth" energy dependence of level density.

\section{Specific of determination of $\rho$ and $k$ from $I_{\gamma\gamma}$} 

The problems in determination of $\rho$ and $k$ and possibility of their 
solution can be most easily analysed on the example well studied nucleus with 
relatively low level density but complicated enough that the level density 
and radiatve strength functions in it can be presented as ``smooth" functions. 
Between the nuclei studied by us, $^{60}Co$ well satisfies these conditions. 
Total intensity $\sum i_1$ of all observed intensities $i_1$ of the cascade 
primary transitions in this nucleus exceeds 76\% and the parameter 
$d=\sum i_\gamma E_\gamma /B_n$= 95\%  [12], respectively. 
This means that the low energy primary transitions are practically absent in 
the data [12]. Because they form continuous component in spectra of the 
two-step cascades, one can  assume that joint analysis of the data on two-step 
cascades and single gamma-transitions will allow appearing of main behavior of
the gamma-decay process of this compound nucleus. In the other words, this 
provides obtaining of rather reliable data on  $\rho$ and $k$ and possibility 
to test them. Total intensity of two-step cascades $I_{\gamma\gamma}$ with the 
sum energy $E_1+E_2=B_n-E_f$ in $^{60}Co$ equals 63.2(9) \% of decays for 
$E_f \leq 1068$ keV. Well established decay scheme allows us to allocate all the 
observed two-step gamma-cascades. This information is necessary  for determination 
of quanta ordering in cascades whose intermediate levels are depopulated by the 
only transition. This is needed for determination [3] of dependence
 $I_{\gamma\gamma}=F(E_1)$ (Fig.~1) and population of the cascade 
 intermediate levels, as well.

Energy dependence of level density derived from these data has clearly 
expressed ``steplike" structure as practically all nuclei studied by the 
method [2].

\section{Reduction of systematic errors}

$I_{\gamma\gamma}$ from Eq.~(1) is determined by values of three unknown 
parameters: total density of the cascade intermediate levels in a given 
energy interval and sums of radiative strength functions of the primary and 
secondary dipole gamma-transitions. Strong anti-correlation of these 
parameters results in rather narrow interval of variations of sums of level 
density with different parity and spin (spin interval is rather unambiguously 
determined by multipolarity selection rule) and sum of strength functions of 
$E1$ and $M1$ transitions which precisely ($\chi^2/f <<1$) reproduce 
$I_{\gamma\gamma}$. It is obvious that interval of variations separately for 
levels with $\pi=+$, $\pi=-$ and $k(E1)$, $k(M1)$ is noticeably wider than 
that for their sums. This statement is true only in case when relation 
between partial widths of primary and secondary transitions is set over whole 
interval of their variations on the grounds of some information. At the lack 
of this information, the only possibility to determine $\rho$ and  $k$ in the 
frameworks of vethod [2] is to take an assumption about equal form of their 
energy dependence. Although partial compensation of this incorrect approach 
in [2] is provided, for example, in case of sign-variable deviation of $k$ for
secondary transitions from that for primary transitions at different 
gamma-transition energies. I.e., relative change in the total radiative width 
$\Gamma_l$ of the cascade intermediate level can be considerably less than 
relative change in $k$.

But even in this case the main specific uncertainty of method [2] results from
 application of notion of equal energy dependence of radiative strength 
 functions for primary and secondary gamma-transitions. In principle, this 
 problem can be solved: it is necessary to measure intensity distribution of 
 cascades to maximum possible number of levels $E_f$ and analyse these data 
 in  appropriate manner. There are no technical obstacles for this variant of 
 determination of $\rho$
 and $k$.

Partial solution of this problem on the base of accumulated information was 
suggested for the first time in [10]. Total population of levels $P=i_1 \times
 i_2/i_{\gamma\gamma}$ equals the product of summed population of all 
 higher-lying levels and branching ration at their decay. Practically the 
 same data on $i$ needed for determination of $P$ for $^{60}Co$ are listed in 
 [11] and [12], intensities of the energy resolved individual cascades
 $i_{\gamma\gamma}$ are given in [15].

As it is seen from Fig.~2, cascade population $P-i_1$ summed over small energy
 interval cannot be reproduced in calculation using existing model ideas of 
 $rho$ [16,17] end $\rho$ [18]. Although the use of $\rho$  and $k$ determined 
 according to [2] gives better result, but complete simultaneous agreement 
 between the experiment and calculation for cascade intensities and population 
 of levels can be achieved only if analysis take into account dependence of 
 $k$ on the energy of decaying level.

In the variant of accounting for different energy dependence of $k$ for 
primary and secondary transitions in form $k^{sec}(E_\gamma, E_{ex})=
k^{prim}(E_\gamma)\times h(E_{ex})$ suggested in [10], the best values for 
$^{60}Co$ slightly differ from that obtained according to [2] (figs.~3 and 4). 
The function $h$ is shown in figs.~3 and 4, as well.

\section{Estimation of significance of observed parameters $\rho$ and $k$}

An ensemble of the experimental data obtained for $^{60}Co$ includes the 
gamma-ray spectrum following thermal neutron radiative capture [19], too. 
This independent information is suitable for estimation of the confidence 
level for the obtained parameters $\rho$  and $k$. The experimental and 
calculated spectra for different values of level density and radiative 
strength functions is performed in Fig.~5.  

It confirms conclusion about inapplicability of models like [16-18] for 
precision description of cascade gamma-decay of heavy nucleus. Besides, 
this points at the necessity of further analysis of the specific of this 
process, decrease and more accurate accounting for influence of all 
sources of systematic errors on the desired parameters. First of all, it 
should be decrease in error of distribution $I_{\gamma\gamma}=F(E_1)$  
and determination of function $k=\phi(E_\gamma, E_{ex})$ directly from 
experiment. Small ``bump" in the region $E_\gamma =3.5$ MeV can be resulted 
from the error cascade intensity  in Fig.~1 and difference of function 
$h(E_{ex})$ from that used in [10].

\section{Conclusion}
Analysis of all totality of information obtained at the thermal neutron 
capture in $^{59}Co$ confirms made earlier for other nuclei conclusion about 
impossibility to achieve agreement between the observed values of functional 
of the cascade gamma-decay process with those calculated in the frameworks of 
existing models of level density and radiative strength functions like 
[16-18] within the limits of experimental error.

In accordance with the model ideas [20], the reason of this phenomenon is 
breaking of at least one Cooper pair of nucleons in this nucleus that is not 
taking into account in other models.\\

  {\bf References} \\\\
1. Zhuravlev B.V., Lychagin A.A., Titarenko N.N., Trykova V.I.,\\
\hspace*{14pt}
 Interaction
of Neutrons with Nuclei: Proc. of XII International Seminar,
Dubna,\\\hspace*{14pt}
 May 2004,
E3-2004-169, Dubna, 2004, 110.\\
2. Vasilieva E.V., Sukhovoj A.M., Khitrov V.A.,
 Phys. At. Nucl. (2001) {\bf 64(2)} 153.
\\ \hspace*{14pt}   Khitrov V. A., Sukhovoj A. M.,
 INDC(CCP)-435, Vienna, 2002, p. 21.\\
\hspace*{14pt} http://arXiv.org/abs/nucl-ex/0110017\\ 
3. Boneva S. T., Khitrov V.A., Sukhovoj A.M.,
 Nucl.  Phys. (1995) {\bf A589}  293.\\ 
4. Bartholomew G.A.  et al.,
Advances in nuclear physics (1973) {\bf 7}  229.\\
5. Voinov A. et all, Phys.\ Rev.,\ C (2001) {\bf 63(4)} 044313-1.\\
6. Khitrov V.A.,  Li Chol, Sukhovoj A.M., 
In: XII International Seminar on Interaction\\\hspace*{14pt}
of Neutrons with Nuclei,  Dubna, May 2004,
E3-2004-169, Dubna, 2004, p. 38.\\
7. Sukhovoj A. M., Khitrov V. A., Li Chol, 
\\\hspace*{14pt}
  JINR Communication E3-2004-100. Dubna, \hspace*{14pt}2004.\\\hspace*{14pt}
Khitrov V.A., Sukhovoj A.M.,  Pham Dinh Khang, Vuong Huu Tan, NguyenXuan Hai,\\\hspace*{14pt}
In: XI International Seminar on Interaction\\\hspace*{14pt}
of Neutrons with Nuclei,
E3-2004-22, Dubna, 2003, p. 107.\\
8. Guttormsen M. at all, Nucl.\ Instrum.\
Methods Phys.\ Res.\ A (1987) \bf 255\rm  518.\\
9. Schiller A. et al., Nucl. Instrum. Methods Phys. Res. (2000)  {\bf A447} 498.\\
10. http://arXiv.org/abs/nucl-ex/0406030; http://arXiv.org/abs/nucl-ex/0410015\\
11. Lone M. A., Leavitt R. A., Harrison  D. A.,
Nuclear Data Tables  (1981) {\bf 26(6)} 511.\\
12. http://www-nds.iaea.org/pgaa/egaf.html\\  
\hspace*{14pt}
 Molnar G. L. et al., App. Rad. Isot. 
 (2000)  {\bf 53} 527.\\
13. Sukhovoj A.M., Khitrov V.A., Phys.Atomic Nuclei (1999) {\bf 62} 19.\\
14. Khitrov V.A.,  Li Chol, Sukhovoj A.M., 
In: XI International Seminar on Interaction\\\hspace*{14pt}
of Neutrons with Nuclei,  Dubna, May 2003,
E3-2004-22, Dubna, 2004, p. 98.\\
15. Sukhovoj A.M., Khitrov V.A., Izv. RAN, ser. fiz., (2003) {\bf 67(5)} 722.\\
16. Kadmenskij S.G., Markushev V.P., Furman W.I., Yad. Fiz. (1983) {\bf 37}
   \hspace*{14pt}277.\\ 
17.  Axel P.,   Phys. Rev.  (1962) {\bf  126(2)}  671. 
\\\hspace*{14pt}  Blatt J. M.,  Weisskopf V. F., Theoretical Nuclear Physics, 
New York (1952).\\
18.  Dilg  W., Schantl W., Vonach  H., Uhl  M.,  Nucl. Phys. (1973) 
{\bf  A217} 269.\\ 
19. Groshev L.V. et al.   Nuc. Data Table  (1968) {\bf 5(1-2)}.\\
20.  Ignatyuk  A.V., Sokolov Yu.V. , Yad. Fiz. (1974)  {\bf 19} 1229.\\
\newpage
\begin{figure}
\vspace{-2cm}
\leavevmode
\epsfxsize=13.0cm

\epsfbox{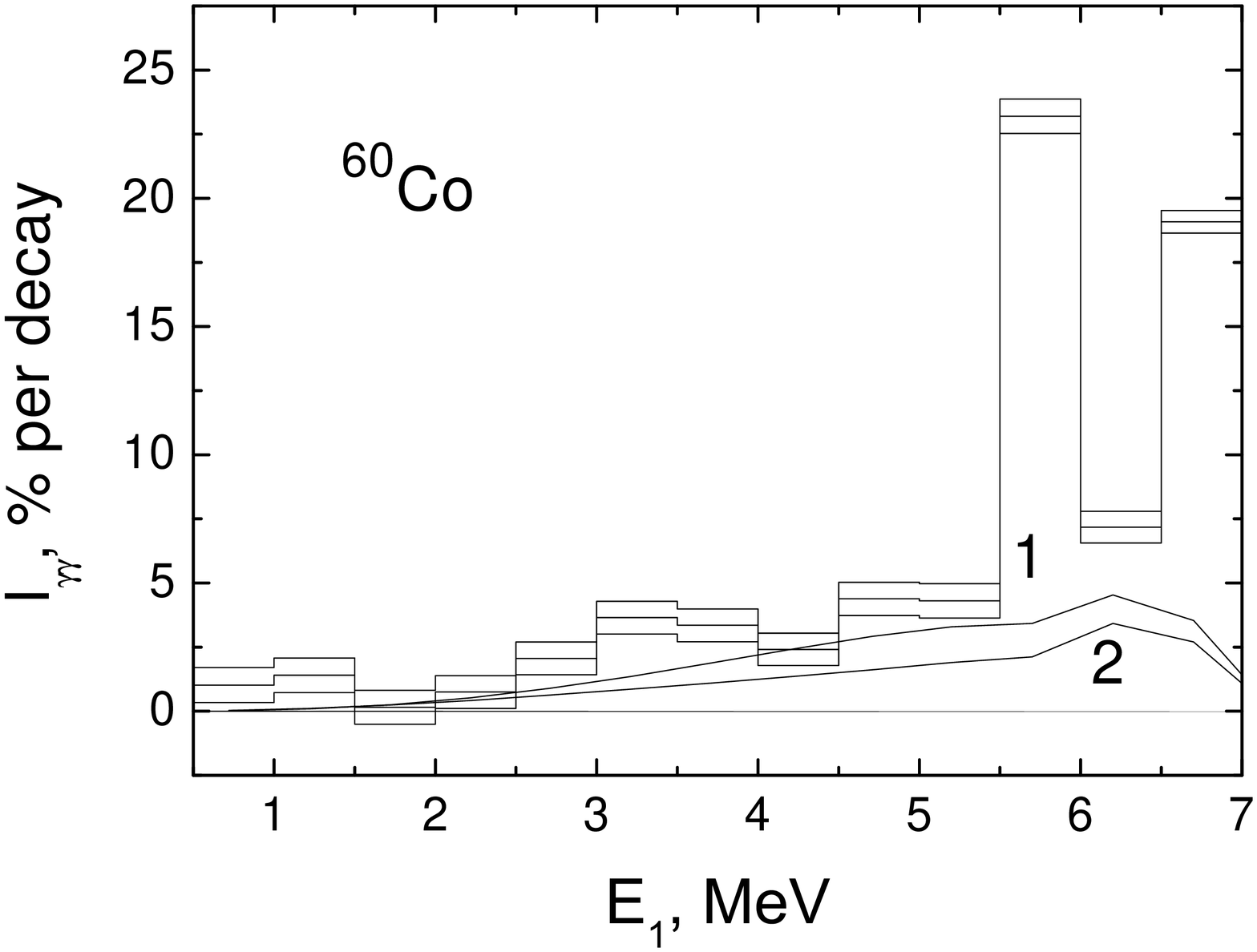}

{\bf  Fig. 1.} 
The total intensity of two-step cascades as a function of the primary 
transition
energy $E_1$ summed over the energy bins of 0.5 MeV for $^{60}$Co. 
The histogram represents experimental data with ordinary statistics errors, 
curves 1 and 2 show calculation by Eq.~1 within models [17,18] and [16,18], 
respectively.
\end{figure}
\begin{figure}
\vspace{-2cm}
\leavevmode
\epsfxsize=13.0cm
\epsfbox{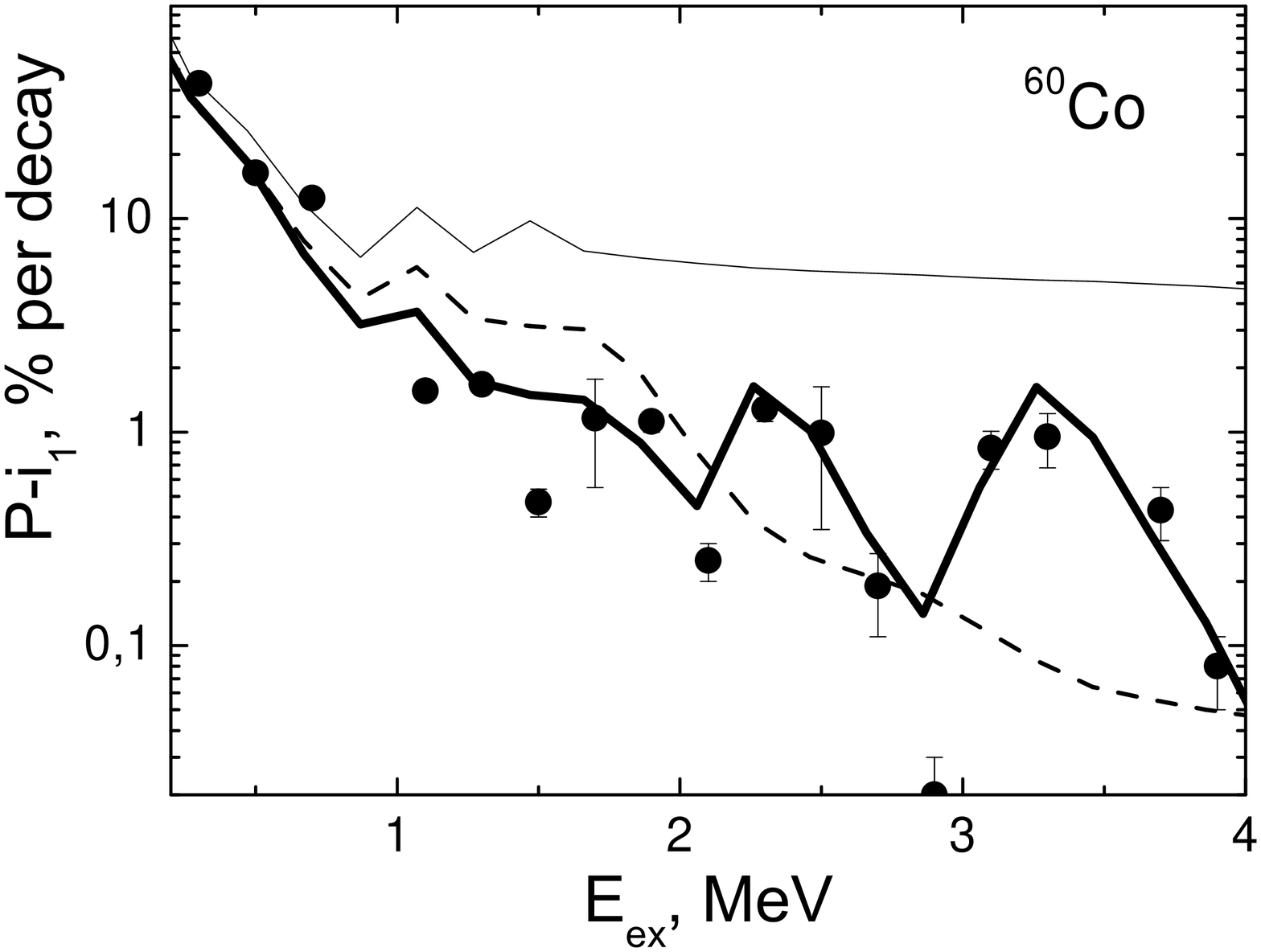}

{\bf  Fig. 2.} 
The total cascade population of levels in the 200 keV energy bins.
Thin line represents calculation within models [17,18], dashed  line shows 
results of calculation using data [2]. Thick line shows results of 
calculation using level density [2], and corresponding strength functions of 
secondary transitions were multiplied by function $h$ [10] presented in 
figs. 3,4.\\
\\
\end{figure}
\newpage
\begin{figure}
\vspace{-2cm}
\leavevmode
\epsfxsize=13.0cm

\epsfbox{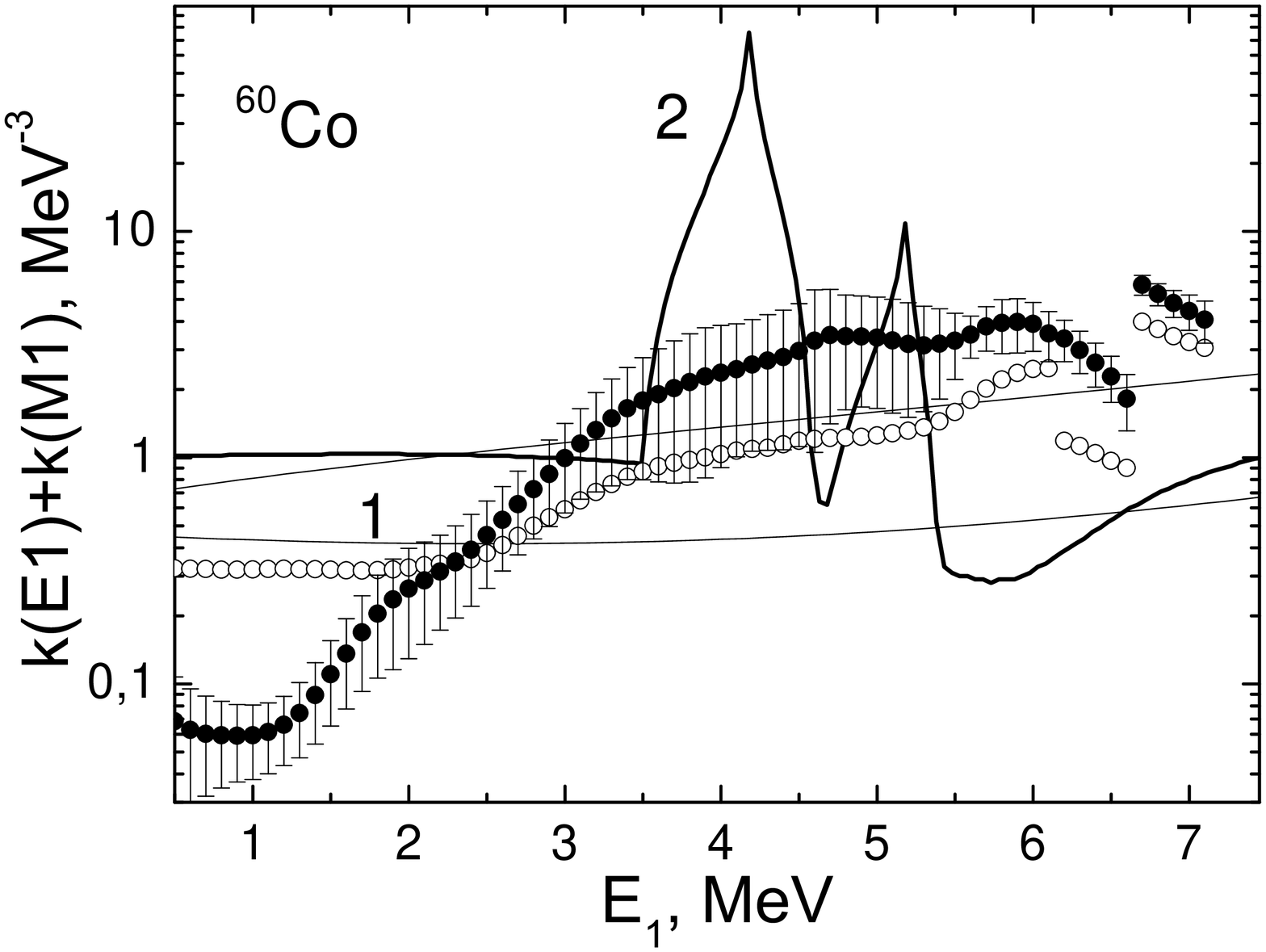}
\epsfxsize=13.0cm

{\bf  Fig. 3.}  The sum of the radiative strength  functions (multiplied by 
$10^9$) for $E1$ and $M1$ transitions (points with errors)  allowing precise 
reproduction of the two-step cascade intensities for $h \neq 1$. Open circles show 
similar values in case $h=1$. Upper and lower thin curves show predictions 
according to models [17] and [16] under assumption $k(M1)$=const, respectively. 
Thick curve shows an example of function $h$ for minimum possible $E_2$.
\end{figure}
\begin{figure}
\leavevmode
\epsfxsize=13.0cm

\epsfbox{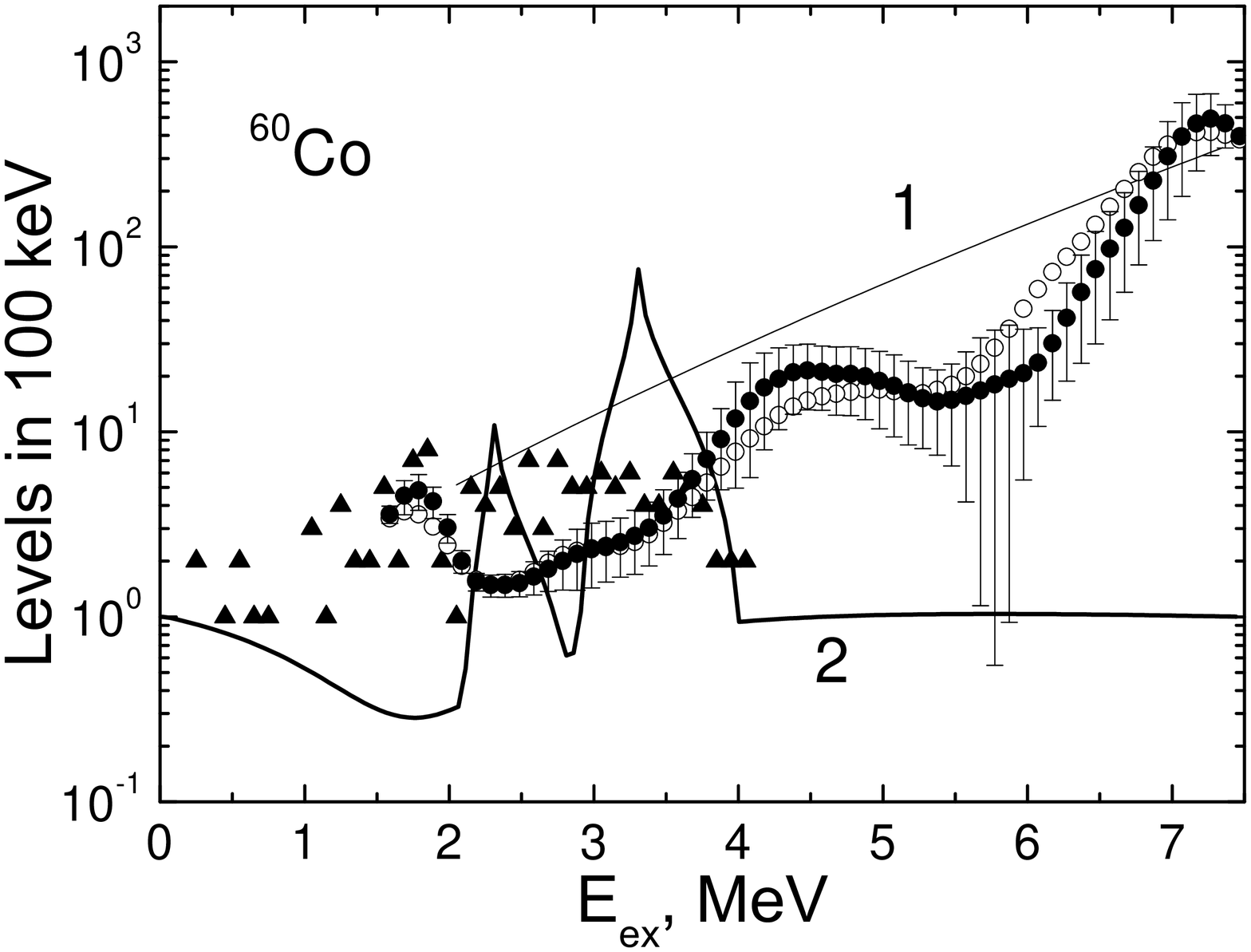}

{\bf  Fig. 4.} The total number of cascade intermediate levels (points with 
errors) allowing  reproduction of the complete set of the experimental data.
Open points show similar values for the case $h=const$. Curve 1 shows 
predictions according model [18]. Curve 2 shows the function $h(E_{ex})$ in case 
$E_2$=0. Triangles demonstrate  observed number of intermediate levels in 
resolved cascades.\\
\end{figure}
\newpage
\begin{figure}
\vspace{-4cm}
\leavevmode
\epsfxsize=13.0cm

\epsfbox{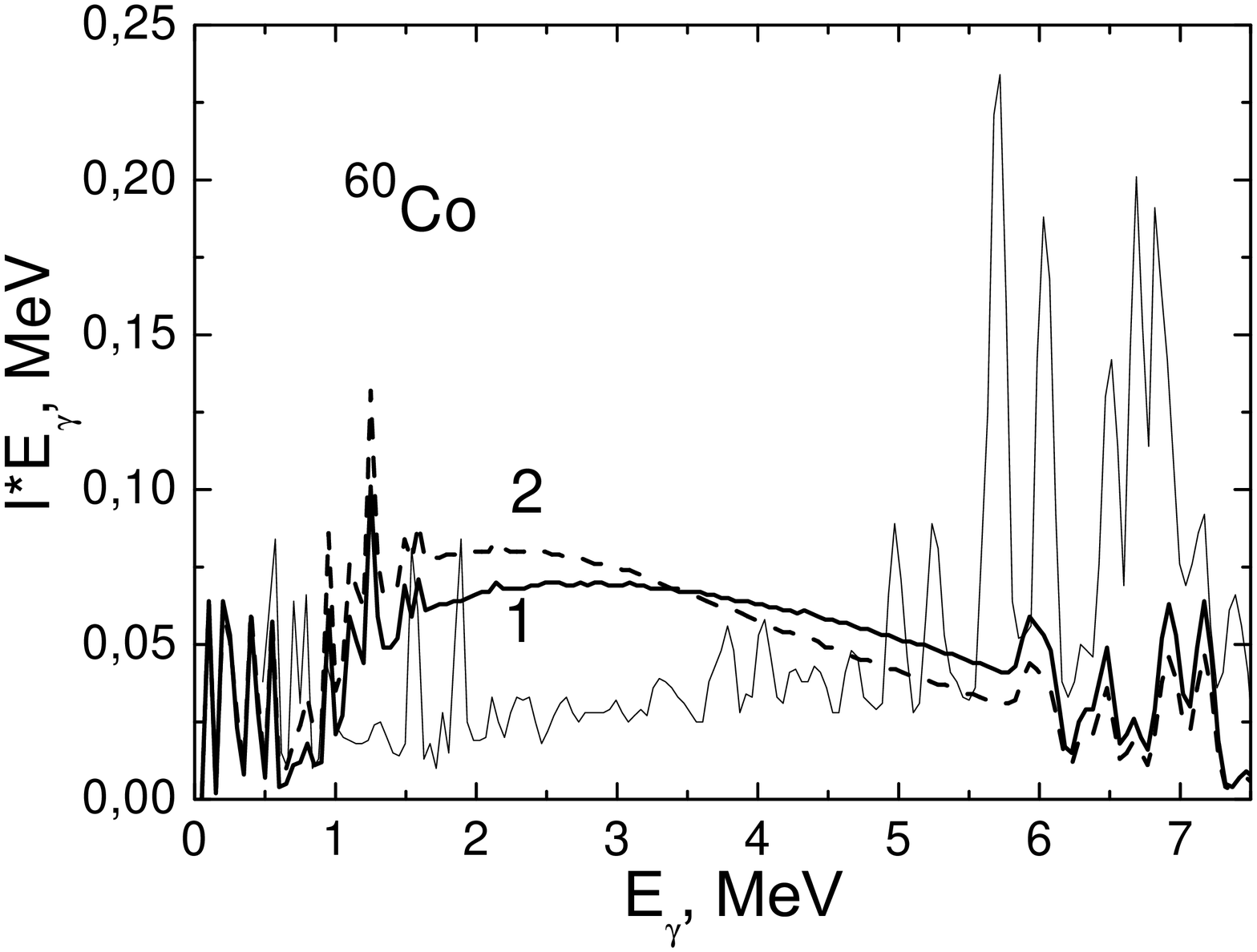}
\vspace{1cm}
\epsfxsize=13.0cm
\epsfbox{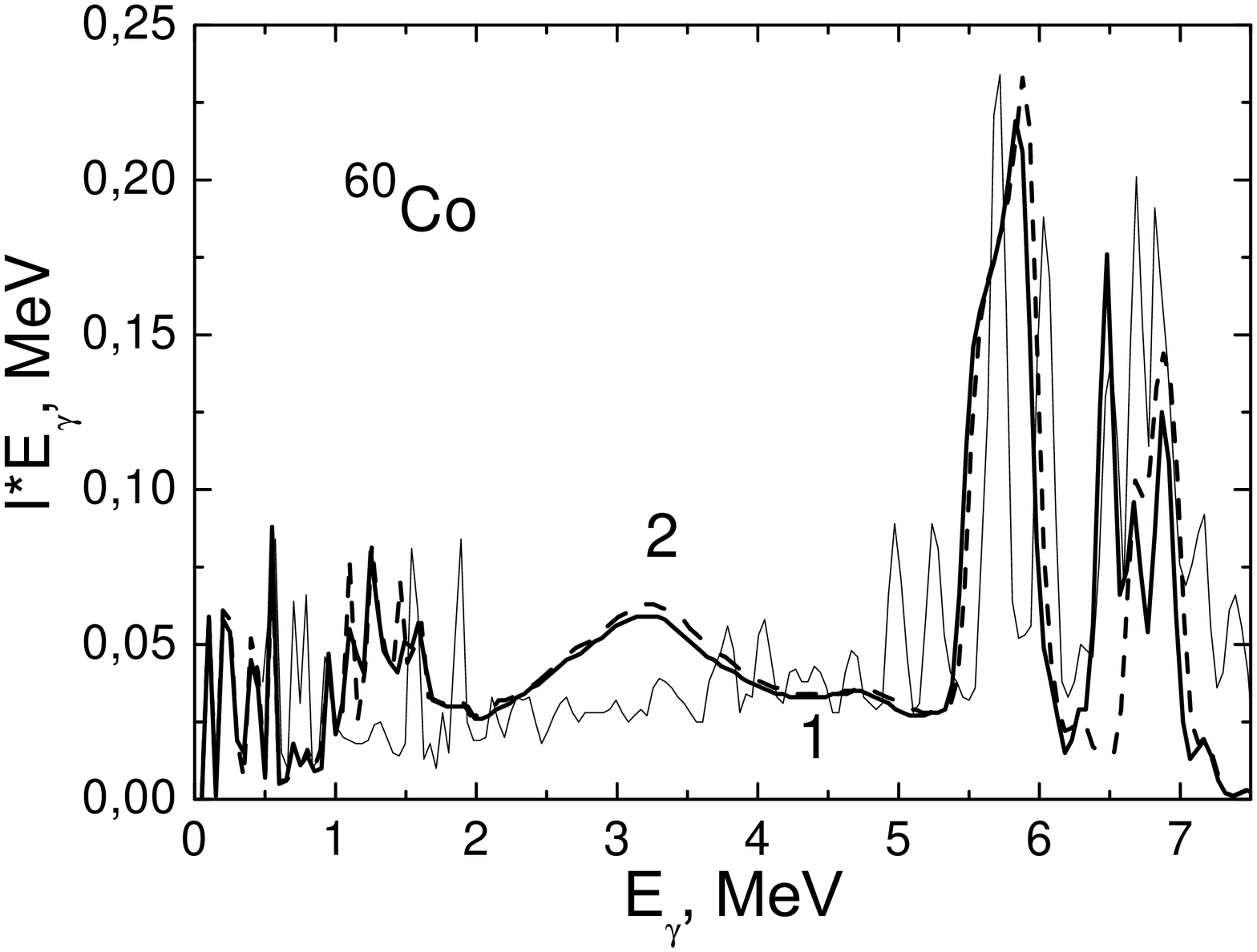}
\vspace{1cm}

{\bf  Fig. 5.} The total gamma-ray spectrum following thermal neutron 
radiative capture. Upper graph: curves 1 and 2 represent results of 
calculation according to models [17,18] and [16,18], respectively. Lower 
graph: curve 1 represents calculation using $\rho$ and $k$ from [2], curve 
2 shows calculation which accounts for different energy dependence of primary 
and secondary transitions.

\end{figure}
\end{document}